\documentclass[%
 reprint,
 amsmath,amssymb,
aps,
]{revtex4-1}

\usepackage{hyperref}
\usepackage{cleveref}
\usepackage{graphicx}% Include figure files
\usepackage{dcolumn}% Align table columns on decimal point
\usepackage{bm}% bold math
\usepackage{epstopdf}
\usepackage{mathtools}
\usepackage{amsmath}
\usepackage{array}
\usepackage{amsmath,amssymb}
\usepackage{lipsum}
\usepackage{subcaption}
\usepackage{CJK}
\usepackage[normalem]{ulem}
\begin{document}

\preprint{APS/123-QED}

\title{Competition Between Intermodal Modulation Instability and Kerr Beam Self-cleaning in Graded-index Multimode Fiber}% Force line breaks with \\

\author{Partha Mondal}
 \altaffiliation[Also at ]{Department of Physics, Indian Institute of Technology, Kharagpur - 721302, India, Email: parthaphotonica@gmail.com.}%Lines break automatically or can be forced with \\
\author{Shailendra K. Varshney}
 \altaffiliation[Also at ]{Department of E $\&$ ECE, Indian Institute of Technology, Kharagpur - 721302, India, Email: skvarshney@ece.iitkgp.ac.in.}%Lines break automatically or can be

\date{\today}% It is always \today, today,

\begin{abstract}
We report the suppression of intermodal modulation instability peaks as a consequence of Kerr induced self beam-cleaning in a 90m long graded-index multimode optical fiber under various specific launching conditions. Output spectrum and the modal beam profiles for three modes ($LP_{01}$, $LP_{11}$ and $LP_{21}$) have been recorded for several values of pump pulse energy. Experimental findings establish that the nonlinear coupling among the guided modes leads to reshaping  the output speckle pattern into a bell shape or higher-order spatially clean beam profile as the pump peak power increases, whereas in spectral domain intermodal modulation instability peaks are observed whose amplitude increases gradually to its maximum value and beyond certain threshold power, the intermodal modulation instability peaks diminish. This suggests the possibility to generate intermodal modulation instability free broadband spectra at high pump peak powers.
\end{abstract}

\maketitle

%\tableofcontents

\section{\label{sec:level1}Introduction}

 Multimode fibers (MMFs) which were overlooked for past many decades, are gaining resurgence as these fibers can overcome the limitations associated with single-mode fibers (SMFs). MMFs effectively address the ever-increasing demand for high data transmission capacities and large power handling capabilities by providing enhanced bandwidth through spatial division multiplexing (SDM) and large core diameter, respectively \cite{richardson2013space,stuart2000dispersive,meng2019multimode}. The large number of modes in MMF interact with each other through several ways and make MMF a perfect natural tool to investigate complex linear and nonlinear spatiotemporal dynamics. In recent years, considerable amount of research has been reported unveiling nonlinear phenomena in MMFs such as multimode soliton dynamics \cite{zhu2016observation,wright2015spatiotemporal,renninger2013optical}, geometric parametric instability \cite{krupa2016observation,teugin2017spatiotemporal}, broadband supercontinuum generation \cite{eftekhar2017versatile,lopez2016visible,teugin2018cascaded,kubat2016multimode}, spatio-temporal mode-locked MMF laser \cite{wright2017spatiotemporal,fu2018several,ding2019multiple,wang2018high} and soliton self-mode conversion \cite{ramachandran2019soliton}. Very recently, experimental investigations on nonlinear propagation in MMFs reveal an exciting nonlinear phenomena named as Kerr-induced beam self-cleaning (KBSC). So far the spatial beam cleaning in MMFs has been effectively realized through nonlinear dissipative processes such as stimulated Raman scattering (SRS) \cite{russell2002stimulated,flusche2006multi,terry2007explanation} or stimulated Brillouin scattering (SBS) \cite{rodgers1999laser,steinhausser2007high}. However, these nonlinear processes do not lead to self-cleaning of the input laser beam and also no Raman beam cleanup can be achieved in step-index MMFs \cite{terry2007explanation}. On the other hand, KBSC leads to reshape the random speckle beam in MMFs at low pump powers into self-cleaned transverse modal profile at higher power levels. The self-cleaning of the transverse output pattern is highly sensitive to the initial launching condition \cite{deliancourt2019kerr}. This nonlinear evolution of spatial beam profile through KBSC process can be attributed to the intermodal four-wave mixing, cross-phase modulation and group delay dispersion effects. Coupling among multiple modes leads to periodic longitudinal modulation of intensity pattern which in turn generate dynamic long-periodic grating exploiting nonlinear Kerr effect. This allows phase-matching processes and promotes the exchange of energy among the transverse modes \cite{schnack2015ultrafast,mondal2018all,hellwig2014experimental}. The incorporation of self-phase modulation in nonlinear energy exchange among the guided modes makes the process non-reciprocal in nature. As a result the energy transmitted into the lower-order transverse mode irreversibly trapped in that particular mode \cite{aschieri2011condensation}. The detailed theoretical analysis addressing the complex mechanism of KBSC is still under investigation. However, the nonlinear reshaping through KBSC can be well reproduced numerically by solving nonlinear Schrodinger equation (NLS) or Gross-Pitaievski equation \cite{krupa2017spatial}.
 There exists three possible mechanisms which play a crucial role in the generation of KBSC process: multimode wave condensation, self-organized instability and irreversible nonlinear mode-coupling process. More recently, KBSC has been explained and experimentally confirmed in the landscape of hydrodynamic 2D turbulence where the mode condensation occurs due to the parametric mode mixing instabilities which in turn generates multiple nonlinear interacting modes with random phases producing an optical wave turbulence \cite{podivilov2019hydrodynamic}. KBSC has been demonstrated under various experimental configurations, involving different fiber types such as GRIN MMFs \cite{liu2016kerr,krupa2019nonlinear,deliancourt2019kerr,wright2016self}, step-index MMF \cite{guenard2017kerr}, microstructure MMF \cite{dupiol2018interplay} and tapered fiber \cite{niang2019spatial}; different pump laser with pulse width varying from femtosecond to nanosecond duration \cite{liu2016kerr,deliancourt2019kerr}; and also normal to anomalous dispersion regimes \cite{leventoux2019multimode,niang2019spatial}. Apart from the conventional fiber, more recently KBSC has also been achieved in double-clad ytterbium doped MMF \cite{guenard2017kerr} and lanthanum aluminum silicate oxide glasses MMF \cite{guenard2019spatial} as well.
 However, the fiber length used so far to obtain KBSC varies from few centimeters to few meters ($<$20m). In our experiments, we have used 90 m long standard graded-index MMF (GI-MMF), which effectively reduces the threshold power to initiate modulational instability process. We have observed a dramatic spectral evolution for sub-nanosecond pump pulses under different input launching conditions which are in contrary to the results observed in \cite{krupa2017spatial}, where KBSC occurs without any spectral distortion. We notice a very interesting phenomena where the increase in input pump peak power generates spectral peaks as a consequence of intermodal modulation instability (IM-MI), intermodal four-wave mixing (IM-FWM) and stimulated Raman scattering (SRS) processes. At very high pump power, input speckle modal profile organizes itself to a well-defined transverse modal profile exploiting KBSC process. The most striking observation is that self-organization through KBSC process leads to suppression of IM-MI peaks for different launching conditions. The paper is organized as follows: In section \ref{exp_set_up}, we provide a brief description of our experimental set-up. Section \ref{exp_result} deals with the experimental result where the work has been performed under three series of experiments. In the first series of the experiments, we report experimental observations for normal incidence on the fiber axis where KBSC leads to reshaping input speckle pattern into $LP_{01}$ mode, whereas in the second and the third series of experiments we report for oblique incidence where input speckle pattern self-organizes through KBSC into $LP_{11}$ and $LP_{21}$ modes, respectively. Finally, the conclusion is made in section \ref{conclusion}.

\section{Experimental Set-up}
\label{exp_set_up}
 \begin{figure}[t]
         \centering
 \includegraphics[width=8.5 cm,height=4 cm]{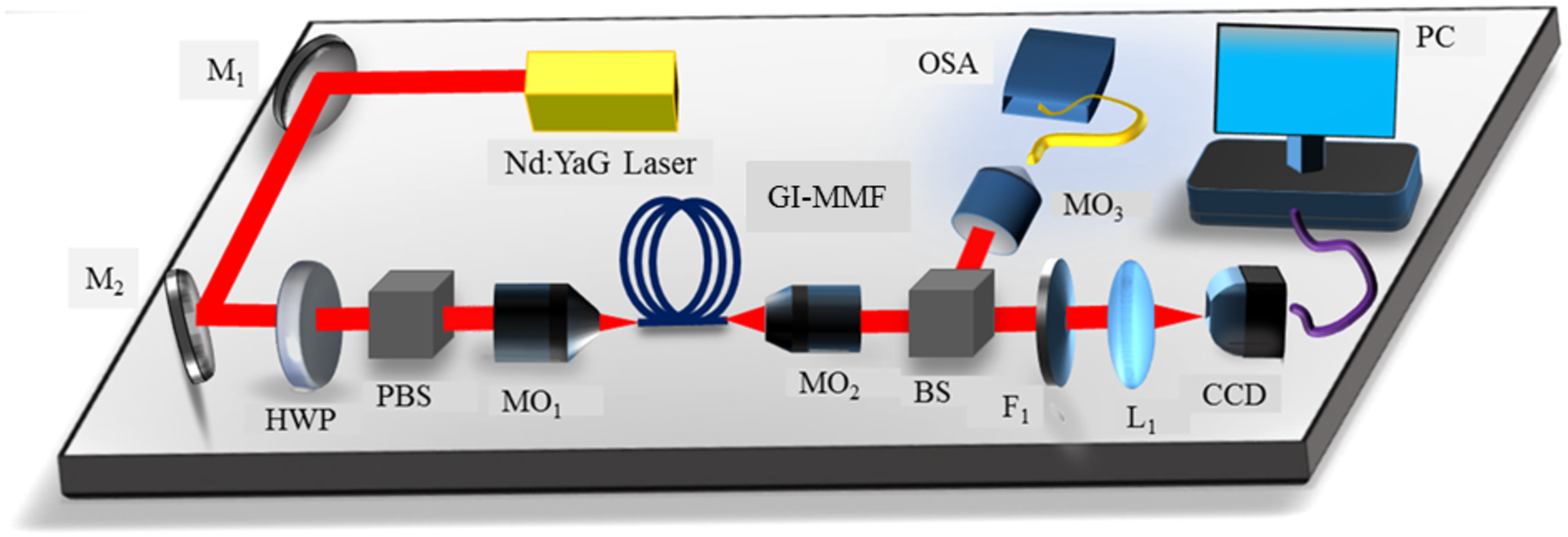}
         \caption{\small{Schematic of the experimental set-up. $M_{1},M_{2}$: silver mirror, HWP: half-wave plate, PBS: polarization beam splitter, $MO_{1},MO_{2},MO_{3}$: microscope objective, BS: plate beam splitter, $F_{1}$ : laser line filter, $L_{1}$: convex lens, CCD: charged coupled device, OSA: optical spectrum analyzer.}}
         \label{Exp_setup}
         \end{figure}

\begin{figure}{}
 \centering
 \begin{subfigure}[b]{0.55\textwidth}
    \includegraphics[width=0.85\linewidth]{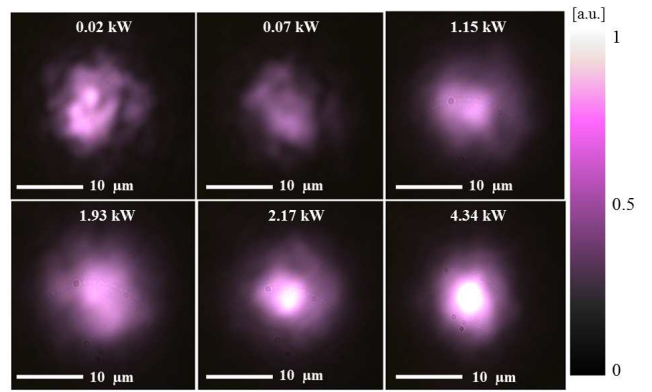}
    \caption{}
    \label{fig:lp01_mode} 
 \end{subfigure}
 
 \begin{subfigure}[b]{0.55\textwidth}
    \includegraphics[width=0.85\linewidth]{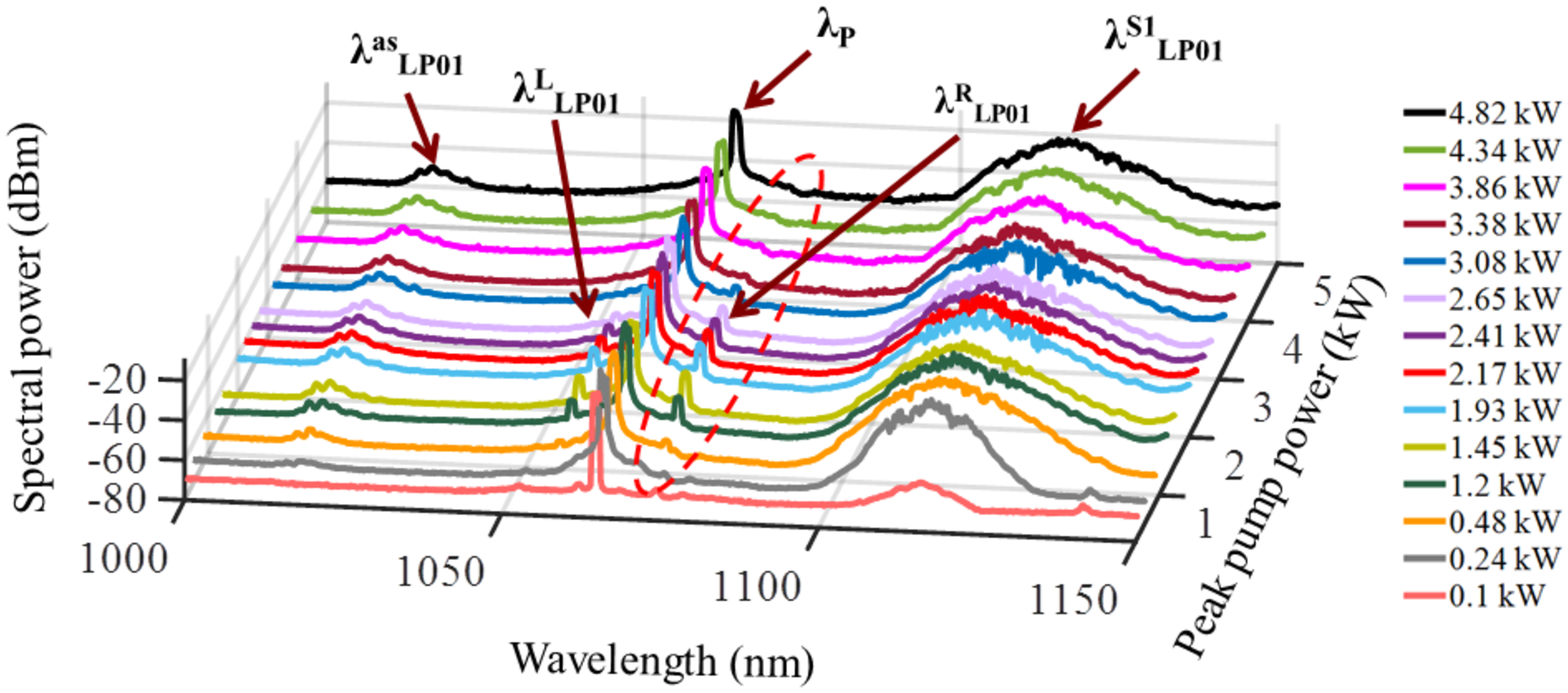}
    \caption{}
    \label{fig:lp01_spectra}
 \end{subfigure}

% \begin{subfigure}[b]{0.55\textwidth}
%    \includegraphics[width=0.7\linewidth]{full_spectral_evolution_LP01_f}
%    \caption{}
%    \label{fig:lp01_full_spectra}
% \end{subfigure}

 \caption{\small{(a) Near-field spatial distributions for normal incidence leading to $LP_{01}$ mode (b) measured output spectra for various pump peak powers.}}
 \end{figure}

  \begin{figure}{}
  \centering
  \begin{subfigure}[b]{0.55\textwidth}
     \includegraphics[width=0.85\linewidth]{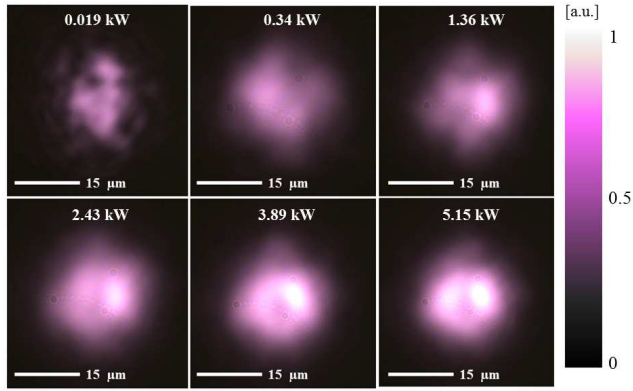}
     \caption{}
     \label{fig:lp11_mode} 
  \end{subfigure}
  
  \begin{subfigure}[b]{0.55\textwidth}
     \includegraphics[width=0.85\linewidth]{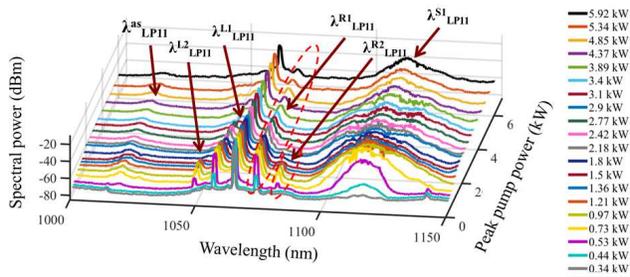}
     \caption{}
     \label{fig:lp11_spectra}
  \end{subfigure}
 
%  \begin{subfigure}[b]{0.55\textwidth}
%     \includegraphics[width=.7\linewidth]{full_spectrum_LP11_mode_f}
%     \caption{}
%     \label{fig:lp11__full_spectra}
%  \end{subfigure}
 
  \caption{\small{(a) Near-field spatial distributions for oblique incidence leading to $LP_{11}$ mode (b) measured output spectra for different pump peak powers.}}
  \end{figure}

  \begin{figure}{}
  \centering
  \begin{subfigure}[b]{0.55\textwidth}
     \includegraphics[width=0.85\linewidth]{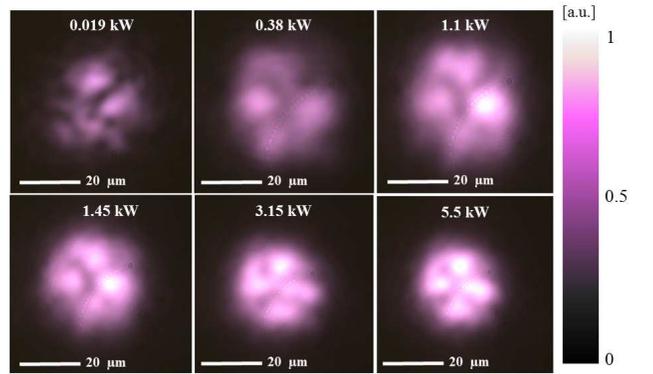}
     \caption{}
     \label{fig:lp21_mode} 
  \end{subfigure}
  
  \begin{subfigure}[b]{0.55\textwidth}
     \includegraphics[width=0.85\linewidth]{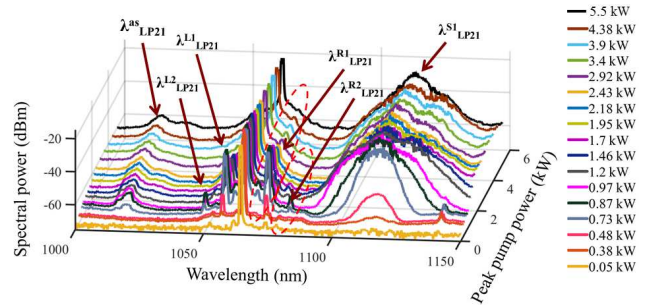}
     \caption{}
     \label{fig:lp21_spectra}
  \end{subfigure}
  
%  \begin{subfigure}[b]{0.55\textwidth}
%       \includegraphics[width=.7\linewidth]{full_spectrum_lp21_f}
%       \caption{}
%       \label{fig:lp21_full_spectra}
%    \end{subfigure}
  \caption{\small{(a) Near-field spatial distributions for oblique incidence leading to $LP_{21}$ mode (b) measured output spectra for several pump peak powers.}}
  \end{figure}
 
 \begin{figure}[t]
          \centering
  \includegraphics[width=8.5 cm,height=4.5 cm]{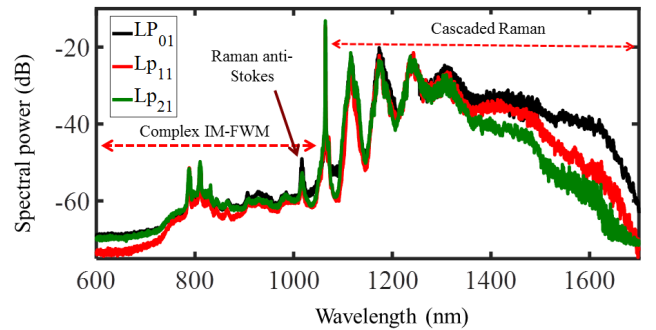}
          \caption{\small{Comparative output spectra for three different spatial modes, $LP_{01}$, $LP_{11}$ and $LP_{21}$ at fixed input pump peak power of 5.25 kW.}}
          \label{full_spectrum_all}
          \end{figure}

 \begin{figure}[t]
          \centering
  \includegraphics[width=8.5 cm,height=4.8 cm]{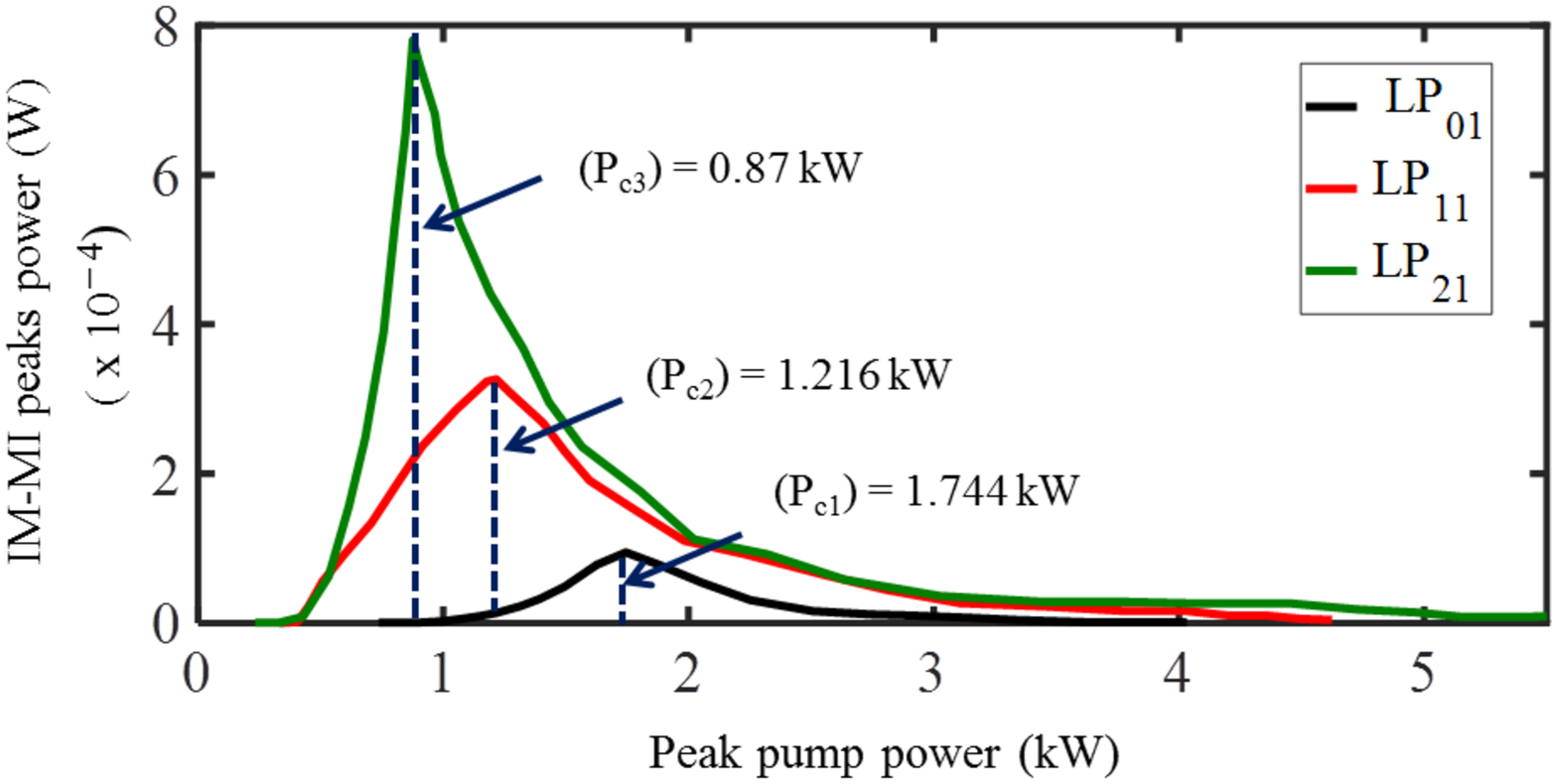}
          \caption{\small{Amplitude variation of IM-MI peaks for different KBSC operations.}}
          \label{MI_peaks_amplitude}
          \end{figure}

The schematic of the experimental setup is shown in Fig. \ref{Exp_setup}. The pump is a single-mode output of a Q-switched microchip Nd:YAG laser source delivering sub-nanosecond pulses ($\sim$ 0.77 ns) with a repetition rate of 23 kHz at 1064 nm. A combination of a polarization beam splitter (PBS) and half-wave plate (HWP) has been employed to control the input pump power coupled to the fiber. The pump is injected into the fiber through a microscope objective (20X, NA 0.4). A 90 m long piece of standard graded-index MMF (GI-MMF) possessing core diameter of 62.5 $\mu$m and numerical aperture 0.275 (Thorlabs GIF625-100) is used. The transverse modal content at the input is controlled using three-axis translational stage. The spatial field distributions have been captured through CMOS ccd camera (digital camera, Thorlabs) while the spectrum is measured by OSA (AQ6319) that has a spectral range from 600 nm to 1700 nm. 

\section{Experimental results}
\label{exp_result}
 
 The experiment has been performed by monitoring the simultaneous evolution of the output spatial distribution and spectrum as a function of power injected to 90m GI-MMF for different launching conditions. We precisely control the excitation of guided modes in the fiber by varying the incident angle.
 
 In the first series of experiments, pump pulses were launched at the normal incidence with the fiber axis. Once the launching condition is fixed, injected pump power is gradually increased by the HWP and PBS. The evolution of output beam shape  at the pump wavelength (1064 nm) with varying input pump powers have been captured through cmos camera employing a laser line filter (FWHM ~3 nm) before the camera. The near-field spatial patterns are shown in Fig. \ref{fig:lp01_mode}. It is observed that at smaller input power values, the output beam profile leads to a speckle pattern which can be attributed to the superposition of large number of excited transverse modes propagating in the fiber with distinct phase velocities. With the increase in pump power, significant transition of the energy distribution takes place towards fundamental mode and the speckle pattern gradually evolve into a bell-shaped beam forming $LP_{01}$ spatial mode. It is observed that threshold value of KBSC to $LP_{01}$ relies around 2.5 kW that suggest a stable output above this threshold value. The experimental results yield self-beam cleaning process to the fundamental mode for on-axis launching condition which has been reported earlier \cite{krupa2019nonlinear,liu2016kerr,deliancourt2019kerr}. The most striking observations are in the spectral domain which is shown in Fig. \ref{fig:lp01_spectra} . Fig. \ref{fig:lp01_spectra} exhibits spectral evolution for various input pump peak powers. It is observed that with the increase in input pump peak power, the output spectrum changes significantly. For relatively low pump peak power (i.e., 0.05 kW), spectral profile exhibits only single peak corresponding to pump wavelength (1064 nm). With the increase in pump power, pair of sidebands at 1056 nm ($\lambda^{R}_{LP01}$) and 1073 nm ($\lambda^{L}_{LP01}$) build up on both side of the pump along with the Raman Stokes at 1117 nm ($\lambda^{S1}_{LP01}$) and an anti-Stoke at 1017 nm ($\lambda^{as}_{LP01}$). The generation of the sidebands can be attributed as intermodal modulation instability (IM-MI) which arises above threshold pump power due to power distribution among different spatial modes following phase-matching condition \cite{arabi2018geometric,guasoni2015generalized}. Threshold value for IM-MI process is observed at ~1 kW of input peak pump power. With the increase in pump power beyond this threshold value, the amplitude of spectral sidebands gradually increase and become maximum at 1.74 kW. Further increase in pump power leads to a decrease in the amplitude of IM-MI peaks and almost disappear above 2.8 kW of input peak pump power.

 In the second series of the experiments, the launching condition of the input beam has been slightly changed, an oblique incidence with respect to the fiber axis is achieved in such a way that maximum power couples into $LP_{11}$ mode. In this alignment, the pump excites a combination of even and odd order of modes where maximum power is coupled into  $LP_{11}$ spatial mode. Fig. \ref{fig:lp11_mode} comprises the evolution of the output beam profile at 1064 nm. At very low pump power (0.02 kW), the superposition of multiple modes lead to a speckle pattern. With the increase in pump power, the output speckle pattern gradually reshapes into $LP_{11}$ mode as shown in the figure. The output images yield that threshold input peak pump power to realize KBSC for $LP_{11}$ mode resemble in the order of 4.8 to 5 kW. The corresponding spectral evolution with varying input peak pump powers is depicted in Fig. \ref{fig:lp11_spectra}. It is observed that threshold peak pump power for the generation of IM-MI peaks is around 0.42 kW. Above this threshold value, IM-MI peaks are generated at 1056 nm ($\lambda^{L1}_{LP11}$)  and 1072 nm ($\lambda^{R1}_{LP11}$). Further increase in pump power generates harmonics at 1048 nm ($\lambda^{L2}_{LP11}$) and 1081 nm ($\lambda^{R2}_{LP11}$) on both sides of the pump which become maximum in amplitude around 1.216 kW. Beyond this input pump power, the amplitude of IM-MI peaks gradually decays and broadband spectrum has been achieved as a consequence of CRS and IM-FWM processes.
  
 In the third and last series of experiments, we demonstrate KBSC for the higher-order $LP_{21}$ mode. To achieve this, the launching angle is tweaked precisely with respect to the fiber axis such that the maximum amount of light couples into $LP_{21}$ mode. After fixing the launching condition, the input power varies by the combination of PBS and HWP. The variation of output beam profiles at 1064 nm for 90 m long GI-MMF is displayed in Fig. \ref{fig:lp21_mode}. At very small pump peak power (0.019 kW), speckle pattern is observed due to superposition of large number of modes propagating through fiber. With the increase in pump power, a transition of speckle pattern into a well defined transverse $LP_{21}$ mode is observed. The peak power threshold value for KBSC in $LP_{21}$ mode is around 5.5 kW which is slightly greater than the KBSC threshold for $LP_{11}$ mode. Corresponding spectral evolution with varying pump power is shown in Fig. \ref{fig:lp21_spectra} which is very much similar as in case of $LP_{11}$ mode. IM-MI peaks are generated for pump peak power $>$0.32 kW and becomes maximum at 0.87 kW. MI peaks are observed at 1056 nm ($\lambda^{L1}_{LP21}$) and 1073 nm ($\lambda^{R1}_{LP21}$), while harmonics are generated at 1048 nm ($\lambda^{L2}_{LP21}$) and 1081 nm ($\lambda^{R2}_{LP21}$) which is almost in the similar spectral positions as observed for KBSC in $LP_{11}$ mode.

 Full spectral evolution is shown in Fig. \ref{full_spectrum_all} which exhibits a complete spectrum for fixed input pump power (5.25 kW) where KBSC reshapes into $LP_{01}$, $LP_{02}$ and $LP_{21}$ spatial modes, respectively. It is observed that for large input pump power, higher-order Raman peaks are generated. At 5.25 kW pump power, higher-order Raman Stokes are generated at 1117 nm, 1173 nm, 1239 nm, 1311 nm, 1445 nm, and 1610 nm, whereas Raman anti-Stoke is generated at 1017 nm. Apart from this, several spectral peaks have been generated in the blue-side of the pump wavelength, similar to self-organization in $LP_{01}$ mode which can be realized through complex IM-FWM process among different mode combinations. The evolution of the amplitude of IM-MI peaks for three different combinations is shown in Fig. \ref{MI_peaks_amplitude}. It is observed that the threshold value to initiate IM-MI process is maximum for the operation where KBSC has been achieved in the fundamental mode and decreases for higher-order mode operation. This is due to the fact that modal combinations of the higher-order modes are more unstable than the fundamental mode which consequently boosts to initiate IM-MI process. Initially, a gradual increment of the amplitude of the IM-MI spectral peaks are observed with an increase in input pump peak power. The maximum amplitude of MI peaks occurs at a peak power of 1.744 kW ($P_{c1}$) for the fundamental mode, whereas it is 1.216 kW ($P_{c2}$) and 0.87 kW ($P_{c3}$) for higher-order modes, $LP_{11}$ and $LP_{21}$, respectively. A further rise in pump peak power reduces the spectral amplitude of IM-MI peaks. This behavior can be realized in the landscape of the KBSC process. With the increase in input pump power, input mode combinations reshape to a well-defined mode which effectively traps the total power in that mode through the KBSC process. Confinement of total power to a particular mode causes the diminishment of the MI gain and finally MI peaks disappear when total power get trapped into a single-mode via the KBSC process. This is due to the fact that MI gain is highly sensitive to the power ratio of the participating modes and it is maximum when power is equally divided among the modes \cite{mondal2019modal}.
  
\section{Conclusion}
\label{conclusion}
To conclude, we have reported experimental investigation of Kerr-induced self-cleaning in a longer length of GI-MMF which reshape transverse speckle pattern into a  spatially clean beam profile. Light injection conditions offer versatile control over the spatial beam cleaning process. Furthermore, we demonstrate the suppression of IM-MI peaks in the spectral domain as a consequence of the beam cleaning process. At high pump peak power, MI free broadband spectrum can be generated  as a consequence of stimulated Raman and IM-FWM processes for every beam-cleaning operations. Our experimental findings on spectral as well as spatial beam cleaning, exploiting Kerr-beam self-cleaning in multimode fiber, will pave the way for developing high power fiber laser and investigation of complex spatio-temporal nonlinear dynamics in multimode platforms.

\bibliography{reference}% Produces the bibliography via BibTeX.

\begin{thebibliography}{40}%
\makeatletter
\providecommand \@ifxundefined [1]{%
 \@ifx{#1\undefined}
}%
\providecommand \@ifnum [1]{%
 \ifnum #1\expandafter \@firstoftwo
 \else \expandafter \@secondoftwo
 \fi
}%
\providecommand \@ifx [1]{%
 \ifx #1\expandafter \@firstoftwo
 \else \expandafter \@secondoftwo
 \fi
}%
\providecommand \natexlab [1]{#1}%
\providecommand \enquote  [1]{``#1''}%
\providecommand \bibnamefont  [1]{#1}%
\providecommand \bibfnamefont [1]{#1}%
\providecommand \citenamefont [1]{#1}%
\providecommand \href@noop [0]{\@secondoftwo}%
\providecommand \href [0]{\begingroup \@sanitize@url \@href}%
\providecommand \@href[1]{\@@startlink{#1}\@@href}%
\providecommand \@@href[1]{\endgroup#1\@@endlink}%
\providecommand \@sanitize@url [0]{\catcode `\\12\catcode `\$12\catcode
  `\&12\catcode `\#12\catcode `\^12\catcode `\_12\catcode `\%12\relax}%
\providecommand \@@startlink[1]{}%
\providecommand \@@endlink[0]{}%
\providecommand \url  [0]{\begingroup\@sanitize@url \@url }%
\providecommand \@url [1]{\endgroup\@href {#1}{\urlprefix }}%
\providecommand \urlprefix  [0]{URL }%
\providecommand \Eprint [0]{\href }%
\providecommand \doibase [0]{http://dx.doi.org/}%
\providecommand \selectlanguage [0]{\@gobble}%
\providecommand \bibinfo  [0]{\@secondoftwo}%
\providecommand \bibfield  [0]{\@secondoftwo}%
\providecommand \translation [1]{[#1]}%
\providecommand \BibitemOpen [0]{}%
\providecommand \bibitemStop [0]{}%
\providecommand \bibitemNoStop [0]{.\EOS\space}%
\providecommand \EOS [0]{\spacefactor3000\relax}%
\providecommand \BibitemShut  [1]{\csname bibitem#1\endcsname}%
\let\auto@bib@innerbib\@empty
%</preamble>
\bibitem [{\citenamefont {Richardson}\ \emph {et~al.}(2013)\citenamefont
  {Richardson}, \citenamefont {Fini},\ and\ \citenamefont
  {Nelson}}]{richardson2013space}%
  \BibitemOpen
  \bibfield  {author} {\bibinfo {author} {\bibfnamefont {D.}~\bibnamefont
  {Richardson}}, \bibinfo {author} {\bibfnamefont {J.}~\bibnamefont {Fini}}, \
  and\ \bibinfo {author} {\bibfnamefont {L.~E.}\ \bibnamefont {Nelson}},\
  }\href@noop {} {\bibfield  {journal} {\bibinfo  {journal} {Nature Photonics}\
  }\textbf {\bibinfo {volume} {7}},\ \bibinfo {pages} {354} (\bibinfo {year}
  {2013})}\BibitemShut {NoStop}%
\bibitem [{\citenamefont {Stuart}(2000)}]{stuart2000dispersive}%
  \BibitemOpen
  \bibfield  {author} {\bibinfo {author} {\bibfnamefont {H.~R.}\ \bibnamefont
  {Stuart}},\ }\href@noop {} {\bibfield  {journal} {\bibinfo  {journal}
  {Science}\ }\textbf {\bibinfo {volume} {289}},\ \bibinfo {pages} {281}
  (\bibinfo {year} {2000})}\BibitemShut {NoStop}%
\bibitem [{\citenamefont {Meng}\ \emph {et~al.}(2019)\citenamefont {Meng},
  \citenamefont {Li}, \citenamefont {Yin}, \citenamefont {Zhang}, \citenamefont
  {Yu}, \citenamefont {Tang}, \citenamefont {Tong},\ and\ \citenamefont
  {Xu}}]{meng2019multimode}%
  \BibitemOpen
  \bibfield  {author} {\bibinfo {author} {\bibfnamefont {Z.}~\bibnamefont
  {Meng}}, \bibinfo {author} {\bibfnamefont {J.}~\bibnamefont {Li}}, \bibinfo
  {author} {\bibfnamefont {C.}~\bibnamefont {Yin}}, \bibinfo {author}
  {\bibfnamefont {T.}~\bibnamefont {Zhang}}, \bibinfo {author} {\bibfnamefont
  {Z.}~\bibnamefont {Yu}}, \bibinfo {author} {\bibfnamefont {M.}~\bibnamefont
  {Tang}}, \bibinfo {author} {\bibfnamefont {W.}~\bibnamefont {Tong}}, \ and\
  \bibinfo {author} {\bibfnamefont {K.}~\bibnamefont {Xu}},\ }\href@noop {}
  {\bibfield  {journal} {\bibinfo  {journal} {AIP Advances}\ }\textbf {\bibinfo
  {volume} {9}},\ \bibinfo {pages} {015004} (\bibinfo {year}
  {2019})}\BibitemShut {NoStop}%
\bibitem [{\citenamefont {Zhu}\ \emph {et~al.}(2016)\citenamefont {Zhu},
  \citenamefont {Wright}, \citenamefont {Christodoulides},\ and\ \citenamefont
  {Wise}}]{zhu2016observation}%
  \BibitemOpen
  \bibfield  {author} {\bibinfo {author} {\bibfnamefont {Z.}~\bibnamefont
  {Zhu}}, \bibinfo {author} {\bibfnamefont {L.~G.}\ \bibnamefont {Wright}},
  \bibinfo {author} {\bibfnamefont {D.~N.}\ \bibnamefont {Christodoulides}}, \
  and\ \bibinfo {author} {\bibfnamefont {F.~W.}\ \bibnamefont {Wise}},\
  }\href@noop {} {\bibfield  {journal} {\bibinfo  {journal} {Optics letters}\
  }\textbf {\bibinfo {volume} {41}},\ \bibinfo {pages} {4819} (\bibinfo {year}
  {2016})}\BibitemShut {NoStop}%
\bibitem [{\citenamefont {Wright}\ \emph {et~al.}(2015)\citenamefont {Wright},
  \citenamefont {Renninger}, \citenamefont {Christodoulides},\ and\
  \citenamefont {Wise}}]{wright2015spatiotemporal}%
  \BibitemOpen
  \bibfield  {author} {\bibinfo {author} {\bibfnamefont {L.~G.}\ \bibnamefont
  {Wright}}, \bibinfo {author} {\bibfnamefont {W.~H.}\ \bibnamefont
  {Renninger}}, \bibinfo {author} {\bibfnamefont {D.~N.}\ \bibnamefont
  {Christodoulides}}, \ and\ \bibinfo {author} {\bibfnamefont {F.~W.}\
  \bibnamefont {Wise}},\ }\href@noop {} {\bibfield  {journal} {\bibinfo
  {journal} {Optics express}\ }\textbf {\bibinfo {volume} {23}},\ \bibinfo
  {pages} {3492} (\bibinfo {year} {2015})}\BibitemShut {NoStop}%
\bibitem [{\citenamefont {Renninger}\ and\ \citenamefont
  {Wise}(2013)}]{renninger2013optical}%
  \BibitemOpen
  \bibfield  {author} {\bibinfo {author} {\bibfnamefont {W.~H.}\ \bibnamefont
  {Renninger}}\ and\ \bibinfo {author} {\bibfnamefont {F.~W.}\ \bibnamefont
  {Wise}},\ }\href@noop {} {\bibfield  {journal} {\bibinfo  {journal} {Nature
  communications}\ }\textbf {\bibinfo {volume} {4}},\ \bibinfo {pages} {1719}
  (\bibinfo {year} {2013})}\BibitemShut {NoStop}%
\bibitem [{\citenamefont {Krupa}\ \emph {et~al.}(2016)\citenamefont {Krupa},
  \citenamefont {Tonello}, \citenamefont {Barth{\'e}l{\'e}my}, \citenamefont
  {Couderc}, \citenamefont {Shalaby}, \citenamefont {Bendahmane}, \citenamefont
  {Millot},\ and\ \citenamefont {Wabnitz}}]{krupa2016observation}%
  \BibitemOpen
  \bibfield  {author} {\bibinfo {author} {\bibfnamefont {K.}~\bibnamefont
  {Krupa}}, \bibinfo {author} {\bibfnamefont {A.}~\bibnamefont {Tonello}},
  \bibinfo {author} {\bibfnamefont {A.}~\bibnamefont {Barth{\'e}l{\'e}my}},
  \bibinfo {author} {\bibfnamefont {V.}~\bibnamefont {Couderc}}, \bibinfo
  {author} {\bibfnamefont {B.~M.}\ \bibnamefont {Shalaby}}, \bibinfo {author}
  {\bibfnamefont {A.}~\bibnamefont {Bendahmane}}, \bibinfo {author}
  {\bibfnamefont {G.}~\bibnamefont {Millot}}, \ and\ \bibinfo {author}
  {\bibfnamefont {S.}~\bibnamefont {Wabnitz}},\ }\href@noop {} {\bibfield
  {journal} {\bibinfo  {journal} {Physical review letters}\ }\textbf {\bibinfo
  {volume} {116}},\ \bibinfo {pages} {183901} (\bibinfo {year}
  {2016})}\BibitemShut {NoStop}%
\bibitem [{\citenamefont {Te{\u{g}}in}\ and\ \citenamefont
  {Orta{\c{c}}}(2017)}]{teugin2017spatiotemporal}%
  \BibitemOpen
  \bibfield  {author} {\bibinfo {author} {\bibfnamefont {U.}~\bibnamefont
  {Te{\u{g}}in}}\ and\ \bibinfo {author} {\bibfnamefont {B.}~\bibnamefont
  {Orta{\c{c}}}},\ }\href@noop {} {\bibfield  {journal} {\bibinfo  {journal}
  {IEEE Photonics Technology Letters}\ }\textbf {\bibinfo {volume} {29}},\
  \bibinfo {pages} {2195} (\bibinfo {year} {2017})}\BibitemShut {NoStop}%
\bibitem [{\citenamefont {Eftekhar}\ \emph {et~al.}(2017)\citenamefont
  {Eftekhar}, \citenamefont {Wright}, \citenamefont {Mills}, \citenamefont
  {Kolesik}, \citenamefont {Correa}, \citenamefont {Wise},\ and\ \citenamefont
  {Christodoulides}}]{eftekhar2017versatile}%
  \BibitemOpen
  \bibfield  {author} {\bibinfo {author} {\bibfnamefont {M.~A.}\ \bibnamefont
  {Eftekhar}}, \bibinfo {author} {\bibfnamefont {L.}~\bibnamefont {Wright}},
  \bibinfo {author} {\bibfnamefont {M.}~\bibnamefont {Mills}}, \bibinfo
  {author} {\bibfnamefont {M.}~\bibnamefont {Kolesik}}, \bibinfo {author}
  {\bibfnamefont {R.~A.}\ \bibnamefont {Correa}}, \bibinfo {author}
  {\bibfnamefont {F.~W.}\ \bibnamefont {Wise}}, \ and\ \bibinfo {author}
  {\bibfnamefont {D.~N.}\ \bibnamefont {Christodoulides}},\ }\href@noop {}
  {\bibfield  {journal} {\bibinfo  {journal} {Optics express}\ }\textbf
  {\bibinfo {volume} {25}},\ \bibinfo {pages} {9078} (\bibinfo {year}
  {2017})}\BibitemShut {NoStop}%
\bibitem [{\citenamefont {Lopez-Galmiche}\ \emph {et~al.}(2016)\citenamefont
  {Lopez-Galmiche}, \citenamefont {Eznaveh}, \citenamefont {Eftekhar},
  \citenamefont {Lopez}, \citenamefont {Wright}, \citenamefont {Wise},
  \citenamefont {Christodoulides},\ and\ \citenamefont
  {Correa}}]{lopez2016visible}%
  \BibitemOpen
  \bibfield  {author} {\bibinfo {author} {\bibfnamefont {G.}~\bibnamefont
  {Lopez-Galmiche}}, \bibinfo {author} {\bibfnamefont {Z.~S.}\ \bibnamefont
  {Eznaveh}}, \bibinfo {author} {\bibfnamefont {M.}~\bibnamefont {Eftekhar}},
  \bibinfo {author} {\bibfnamefont {J.~A.}\ \bibnamefont {Lopez}}, \bibinfo
  {author} {\bibfnamefont {L.}~\bibnamefont {Wright}}, \bibinfo {author}
  {\bibfnamefont {F.}~\bibnamefont {Wise}}, \bibinfo {author} {\bibfnamefont
  {D.}~\bibnamefont {Christodoulides}}, \ and\ \bibinfo {author} {\bibfnamefont
  {R.~A.}\ \bibnamefont {Correa}},\ }\href@noop {} {\bibfield  {journal}
  {\bibinfo  {journal} {Optics letters}\ }\textbf {\bibinfo {volume} {41}},\
  \bibinfo {pages} {2553} (\bibinfo {year} {2016})}\BibitemShut {NoStop}%
\bibitem [{\citenamefont {Te{\u{g}}in}\ and\ \citenamefont
  {Orta{\c{c}}}(2018)}]{teugin2018cascaded}%
  \BibitemOpen
  \bibfield  {author} {\bibinfo {author} {\bibfnamefont {U.}~\bibnamefont
  {Te{\u{g}}in}}\ and\ \bibinfo {author} {\bibfnamefont {B.}~\bibnamefont
  {Orta{\c{c}}}},\ }\href@noop {} {\bibfield  {journal} {\bibinfo  {journal}
  {Scientific reports}\ }\textbf {\bibinfo {volume} {8}},\ \bibinfo {pages}
  {12470} (\bibinfo {year} {2018})}\BibitemShut {NoStop}%
\bibitem [{\citenamefont {Kubat}\ and\ \citenamefont
  {Bang}(2016)}]{kubat2016multimode}%
  \BibitemOpen
  \bibfield  {author} {\bibinfo {author} {\bibfnamefont {I.}~\bibnamefont
  {Kubat}}\ and\ \bibinfo {author} {\bibfnamefont {O.}~\bibnamefont {Bang}},\
  }\href@noop {} {\bibfield  {journal} {\bibinfo  {journal} {Optics express}\
  }\textbf {\bibinfo {volume} {24}},\ \bibinfo {pages} {2513} (\bibinfo {year}
  {2016})}\BibitemShut {NoStop}%
\bibitem [{\citenamefont {Wright}\ \emph {et~al.}(2017)\citenamefont {Wright},
  \citenamefont {Christodoulides},\ and\ \citenamefont
  {Wise}}]{wright2017spatiotemporal}%
  \BibitemOpen
  \bibfield  {author} {\bibinfo {author} {\bibfnamefont {L.~G.}\ \bibnamefont
  {Wright}}, \bibinfo {author} {\bibfnamefont {D.~N.}\ \bibnamefont
  {Christodoulides}}, \ and\ \bibinfo {author} {\bibfnamefont {F.~W.}\
  \bibnamefont {Wise}},\ }\href@noop {} {\bibfield  {journal} {\bibinfo
  {journal} {Science}\ }\textbf {\bibinfo {volume} {358}},\ \bibinfo {pages}
  {94} (\bibinfo {year} {2017})}\BibitemShut {NoStop}%
\bibitem [{\citenamefont {Fu}\ \emph {et~al.}(2018)\citenamefont {Fu},
  \citenamefont {Wright}, \citenamefont {Sidorenko}, \citenamefont {Backus},\
  and\ \citenamefont {Wise}}]{fu2018several}%
  \BibitemOpen
  \bibfield  {author} {\bibinfo {author} {\bibfnamefont {W.}~\bibnamefont
  {Fu}}, \bibinfo {author} {\bibfnamefont {L.~G.}\ \bibnamefont {Wright}},
  \bibinfo {author} {\bibfnamefont {P.}~\bibnamefont {Sidorenko}}, \bibinfo
  {author} {\bibfnamefont {S.}~\bibnamefont {Backus}}, \ and\ \bibinfo {author}
  {\bibfnamefont {F.~W.}\ \bibnamefont {Wise}},\ }\href@noop {} {\bibfield
  {journal} {\bibinfo  {journal} {Optics express}\ }\textbf {\bibinfo {volume}
  {26}},\ \bibinfo {pages} {9432} (\bibinfo {year} {2018})}\BibitemShut
  {NoStop}%
\bibitem [{\citenamefont {Ding}\ \emph {et~al.}(2019)\citenamefont {Ding},
  \citenamefont {Xiao}, \citenamefont {Wang},\ and\ \citenamefont
  {Yang}}]{ding2019multiple}%
  \BibitemOpen
  \bibfield  {author} {\bibinfo {author} {\bibfnamefont {Y.}~\bibnamefont
  {Ding}}, \bibinfo {author} {\bibfnamefont {X.}~\bibnamefont {Xiao}}, \bibinfo
  {author} {\bibfnamefont {P.}~\bibnamefont {Wang}}, \ and\ \bibinfo {author}
  {\bibfnamefont {C.}~\bibnamefont {Yang}},\ }\href@noop {} {\bibfield
  {journal} {\bibinfo  {journal} {Optics express}\ }\textbf {\bibinfo {volume}
  {27}},\ \bibinfo {pages} {11435} (\bibinfo {year} {2019})}\BibitemShut
  {NoStop}%
\bibitem [{\citenamefont {Wang}\ \emph {et~al.}(2018)\citenamefont {Wang},
  \citenamefont {Tang}, \citenamefont {Yan},\ and\ \citenamefont
  {Xu}}]{wang2018high}%
  \BibitemOpen
  \bibfield  {author} {\bibinfo {author} {\bibfnamefont {Y.}~\bibnamefont
  {Wang}}, \bibinfo {author} {\bibfnamefont {Y.}~\bibnamefont {Tang}}, \bibinfo
  {author} {\bibfnamefont {S.}~\bibnamefont {Yan}}, \ and\ \bibinfo {author}
  {\bibfnamefont {J.}~\bibnamefont {Xu}},\ }\href@noop {} {\bibfield  {journal}
  {\bibinfo  {journal} {Laser Physics Letters}\ }\textbf {\bibinfo {volume}
  {15}},\ \bibinfo {pages} {085101} (\bibinfo {year} {2018})}\BibitemShut
  {NoStop}%
\bibitem [{\citenamefont {Ramachandran}\ and\ \citenamefont
  {Agrawal}(2019)}]{ramachandran2019soliton}%
  \BibitemOpen
  \bibfield  {author} {\bibinfo {author} {\bibfnamefont {S.}~\bibnamefont
  {Ramachandran}}\ and\ \bibinfo {author} {\bibfnamefont {G.~P.}\ \bibnamefont
  {Agrawal}},\ }in\ \href@noop {} {\emph {\bibinfo {booktitle} {Fiber Lasers
  XVI: Technology and Systems}}},\ Vol.\ \bibinfo {volume} {10897}\ (\bibinfo
  {organization} {International Society for Optics and Photonics},\ \bibinfo
  {year} {2019})\ p.\ \bibinfo {pages} {108971T}\BibitemShut {NoStop}%
\bibitem [{\citenamefont {Russell}\ \emph {et~al.}(2002)\citenamefont
  {Russell}, \citenamefont {Willis}, \citenamefont {Crookston},\ and\
  \citenamefont {Roh}}]{russell2002stimulated}%
  \BibitemOpen
  \bibfield  {author} {\bibinfo {author} {\bibfnamefont {T.~H.}\ \bibnamefont
  {Russell}}, \bibinfo {author} {\bibfnamefont {S.~M.}\ \bibnamefont {Willis}},
  \bibinfo {author} {\bibfnamefont {M.~B.}\ \bibnamefont {Crookston}}, \ and\
  \bibinfo {author} {\bibfnamefont {W.~B.}\ \bibnamefont {Roh}},\ }\href@noop
  {} {\bibfield  {journal} {\bibinfo  {journal} {Journal of Nonlinear Optical
  Physics \& Materials}\ }\textbf {\bibinfo {volume} {11}},\ \bibinfo {pages}
  {303} (\bibinfo {year} {2002})}\BibitemShut {NoStop}%
\bibitem [{\citenamefont {Flusche}\ \emph {et~al.}(2006)\citenamefont
  {Flusche}, \citenamefont {Alley}, \citenamefont {Russell},\ and\
  \citenamefont {Roh}}]{flusche2006multi}%
  \BibitemOpen
  \bibfield  {author} {\bibinfo {author} {\bibfnamefont {B.~M.}\ \bibnamefont
  {Flusche}}, \bibinfo {author} {\bibfnamefont {T.~G.}\ \bibnamefont {Alley}},
  \bibinfo {author} {\bibfnamefont {T.~H.}\ \bibnamefont {Russell}}, \ and\
  \bibinfo {author} {\bibfnamefont {W.~B.}\ \bibnamefont {Roh}},\ }\href@noop
  {} {\bibfield  {journal} {\bibinfo  {journal} {Optics express}\ }\textbf
  {\bibinfo {volume} {14}},\ \bibinfo {pages} {11748} (\bibinfo {year}
  {2006})}\BibitemShut {NoStop}%
\bibitem [{\citenamefont {Terry}\ \emph {et~al.}(2007)\citenamefont {Terry},
  \citenamefont {Alley},\ and\ \citenamefont {Russell}}]{terry2007explanation}%
  \BibitemOpen
  \bibfield  {author} {\bibinfo {author} {\bibfnamefont {N.~B.}\ \bibnamefont
  {Terry}}, \bibinfo {author} {\bibfnamefont {T.~G.}\ \bibnamefont {Alley}}, \
  and\ \bibinfo {author} {\bibfnamefont {T.~H.}\ \bibnamefont {Russell}},\
  }\href@noop {} {\bibfield  {journal} {\bibinfo  {journal} {Optics express}\
  }\textbf {\bibinfo {volume} {15}},\ \bibinfo {pages} {17509} (\bibinfo {year}
  {2007})}\BibitemShut {NoStop}%
\bibitem [{\citenamefont {Rodgers}\ \emph {et~al.}(1999)\citenamefont
  {Rodgers}, \citenamefont {Russell},\ and\ \citenamefont
  {Roh}}]{rodgers1999laser}%
  \BibitemOpen
  \bibfield  {author} {\bibinfo {author} {\bibfnamefont {B.~C.}\ \bibnamefont
  {Rodgers}}, \bibinfo {author} {\bibfnamefont {T.~H.}\ \bibnamefont
  {Russell}}, \ and\ \bibinfo {author} {\bibfnamefont {W.~B.}\ \bibnamefont
  {Roh}},\ }\href@noop {} {\bibfield  {journal} {\bibinfo  {journal} {Optics
  letters}\ }\textbf {\bibinfo {volume} {24}},\ \bibinfo {pages} {1124}
  (\bibinfo {year} {1999})}\BibitemShut {NoStop}%
\bibitem [{\citenamefont {Steinhausser}\ \emph {et~al.}(2007)\citenamefont
  {Steinhausser}, \citenamefont {Brignon}, \citenamefont {Lallier},
  \citenamefont {Huignard},\ and\ \citenamefont
  {Georges}}]{steinhausser2007high}%
  \BibitemOpen
  \bibfield  {author} {\bibinfo {author} {\bibfnamefont {B.}~\bibnamefont
  {Steinhausser}}, \bibinfo {author} {\bibfnamefont {A.}~\bibnamefont
  {Brignon}}, \bibinfo {author} {\bibfnamefont {E.}~\bibnamefont {Lallier}},
  \bibinfo {author} {\bibfnamefont {J.-P.}\ \bibnamefont {Huignard}}, \ and\
  \bibinfo {author} {\bibfnamefont {P.}~\bibnamefont {Georges}},\ }\href@noop
  {} {\bibfield  {journal} {\bibinfo  {journal} {Optics Express}\ }\textbf
  {\bibinfo {volume} {15}},\ \bibinfo {pages} {6464} (\bibinfo {year}
  {2007})}\BibitemShut {NoStop}%
\bibitem [{\citenamefont {Deliancourt}\ \emph {et~al.}(2019)\citenamefont
  {Deliancourt}, \citenamefont {Fabert}, \citenamefont {Tonello}, \citenamefont
  {Krupa}, \citenamefont {Desfarges-Berthelemot}, \citenamefont {Kermene},
  \citenamefont {Millot}, \citenamefont {Barth{\'e}l{\'e}my}, \citenamefont
  {Wabnitz},\ and\ \citenamefont {Couderc}}]{deliancourt2019kerr}%
  \BibitemOpen
  \bibfield  {author} {\bibinfo {author} {\bibfnamefont {E.}~\bibnamefont
  {Deliancourt}}, \bibinfo {author} {\bibfnamefont {M.}~\bibnamefont {Fabert}},
  \bibinfo {author} {\bibfnamefont {A.}~\bibnamefont {Tonello}}, \bibinfo
  {author} {\bibfnamefont {K.}~\bibnamefont {Krupa}}, \bibinfo {author}
  {\bibfnamefont {A.}~\bibnamefont {Desfarges-Berthelemot}}, \bibinfo {author}
  {\bibfnamefont {V.}~\bibnamefont {Kermene}}, \bibinfo {author} {\bibfnamefont
  {G.}~\bibnamefont {Millot}}, \bibinfo {author} {\bibfnamefont
  {A.}~\bibnamefont {Barth{\'e}l{\'e}my}}, \bibinfo {author} {\bibfnamefont
  {S.}~\bibnamefont {Wabnitz}}, \ and\ \bibinfo {author} {\bibfnamefont
  {V.}~\bibnamefont {Couderc}},\ }\href@noop {} {\bibfield  {journal} {\bibinfo
   {journal} {OSA Continuum}\ }\textbf {\bibinfo {volume} {2}},\ \bibinfo
  {pages} {1089} (\bibinfo {year} {2019})}\BibitemShut {NoStop}%
\bibitem [{\citenamefont {Schnack}\ \emph {et~al.}(2015)\citenamefont
  {Schnack}, \citenamefont {Hellwig}, \citenamefont {Brinkmann},\ and\
  \citenamefont {Fallnich}}]{schnack2015ultrafast}%
  \BibitemOpen
  \bibfield  {author} {\bibinfo {author} {\bibfnamefont {M.}~\bibnamefont
  {Schnack}}, \bibinfo {author} {\bibfnamefont {T.}~\bibnamefont {Hellwig}},
  \bibinfo {author} {\bibfnamefont {M.}~\bibnamefont {Brinkmann}}, \ and\
  \bibinfo {author} {\bibfnamefont {C.}~\bibnamefont {Fallnich}},\ }\href@noop
  {} {\bibfield  {journal} {\bibinfo  {journal} {Optics letters}\ }\textbf
  {\bibinfo {volume} {40}},\ \bibinfo {pages} {4675} (\bibinfo {year}
  {2015})}\BibitemShut {NoStop}%
\bibitem [{\citenamefont {Mondal}\ \emph {et~al.}(2018)\citenamefont {Mondal},
  \citenamefont {Haldar}, \citenamefont {Mishra},\ and\ \citenamefont
  {Varshney}}]{mondal2018all}%
  \BibitemOpen
  \bibfield  {author} {\bibinfo {author} {\bibfnamefont {P.}~\bibnamefont
  {Mondal}}, \bibinfo {author} {\bibfnamefont {R.}~\bibnamefont {Haldar}},
  \bibinfo {author} {\bibfnamefont {V.}~\bibnamefont {Mishra}}, \ and\ \bibinfo
  {author} {\bibfnamefont {S.~K.}\ \bibnamefont {Varshney}},\ }\href@noop {}
  {\bibfield  {journal} {\bibinfo  {journal} {IEEE Photonics Technology
  Letters}\ }\textbf {\bibinfo {volume} {30}},\ \bibinfo {pages} {2175}
  (\bibinfo {year} {2018})}\BibitemShut {NoStop}%
\bibitem [{\citenamefont {Hellwig}\ \emph {et~al.}(2014)\citenamefont
  {Hellwig}, \citenamefont {Schnack}, \citenamefont {Walbaum}, \citenamefont
  {Dobner},\ and\ \citenamefont {Fallnich}}]{hellwig2014experimental}%
  \BibitemOpen
  \bibfield  {author} {\bibinfo {author} {\bibfnamefont {T.}~\bibnamefont
  {Hellwig}}, \bibinfo {author} {\bibfnamefont {M.}~\bibnamefont {Schnack}},
  \bibinfo {author} {\bibfnamefont {T.}~\bibnamefont {Walbaum}}, \bibinfo
  {author} {\bibfnamefont {S.}~\bibnamefont {Dobner}}, \ and\ \bibinfo {author}
  {\bibfnamefont {C.}~\bibnamefont {Fallnich}},\ }\href@noop {} {\bibfield
  {journal} {\bibinfo  {journal} {Optics express}\ }\textbf {\bibinfo {volume}
  {22}},\ \bibinfo {pages} {24951} (\bibinfo {year} {2014})}\BibitemShut
  {NoStop}%
\bibitem [{\citenamefont {Aschieri}\ \emph {et~al.}(2011)\citenamefont
  {Aschieri}, \citenamefont {Garnier}, \citenamefont {Michel}, \citenamefont
  {Doya},\ and\ \citenamefont {Picozzi}}]{aschieri2011condensation}%
  \BibitemOpen
  \bibfield  {author} {\bibinfo {author} {\bibfnamefont {P.}~\bibnamefont
  {Aschieri}}, \bibinfo {author} {\bibfnamefont {J.}~\bibnamefont {Garnier}},
  \bibinfo {author} {\bibfnamefont {C.}~\bibnamefont {Michel}}, \bibinfo
  {author} {\bibfnamefont {V.}~\bibnamefont {Doya}}, \ and\ \bibinfo {author}
  {\bibfnamefont {A.}~\bibnamefont {Picozzi}},\ }\href@noop {} {\bibfield
  {journal} {\bibinfo  {journal} {Physical Review A}\ }\textbf {\bibinfo
  {volume} {83}},\ \bibinfo {pages} {033838} (\bibinfo {year}
  {2011})}\BibitemShut {NoStop}%
\bibitem [{\citenamefont {Krupa}\ \emph {et~al.}(2017)\citenamefont {Krupa},
  \citenamefont {Tonello}, \citenamefont {Shalaby}, \citenamefont {Fabert},
  \citenamefont {Barth{\'e}l{\'e}my}, \citenamefont {Millot}, \citenamefont
  {Wabnitz},\ and\ \citenamefont {Couderc}}]{krupa2017spatial}%
  \BibitemOpen
  \bibfield  {author} {\bibinfo {author} {\bibfnamefont {K.}~\bibnamefont
  {Krupa}}, \bibinfo {author} {\bibfnamefont {A.}~\bibnamefont {Tonello}},
  \bibinfo {author} {\bibfnamefont {B.~M.}\ \bibnamefont {Shalaby}}, \bibinfo
  {author} {\bibfnamefont {M.}~\bibnamefont {Fabert}}, \bibinfo {author}
  {\bibfnamefont {A.}~\bibnamefont {Barth{\'e}l{\'e}my}}, \bibinfo {author}
  {\bibfnamefont {G.}~\bibnamefont {Millot}}, \bibinfo {author} {\bibfnamefont
  {S.}~\bibnamefont {Wabnitz}}, \ and\ \bibinfo {author} {\bibfnamefont
  {V.}~\bibnamefont {Couderc}},\ }\href@noop {} {\bibfield  {journal} {\bibinfo
   {journal} {Nature Photonics}\ }\textbf {\bibinfo {volume} {11}},\ \bibinfo
  {pages} {237} (\bibinfo {year} {2017})}\BibitemShut {NoStop}%
\bibitem [{\citenamefont {Podivilov}\ \emph {et~al.}(2019)\citenamefont
  {Podivilov}, \citenamefont {Kharenko}, \citenamefont {Gonta}, \citenamefont
  {Krupa}, \citenamefont {Sidelnikov}, \citenamefont {Turitsyn}, \citenamefont
  {Fedoruk}, \citenamefont {Babin},\ and\ \citenamefont
  {Wabnitz}}]{podivilov2019hydrodynamic}%
  \BibitemOpen
  \bibfield  {author} {\bibinfo {author} {\bibfnamefont {E.}~\bibnamefont
  {Podivilov}}, \bibinfo {author} {\bibfnamefont {D.}~\bibnamefont {Kharenko}},
  \bibinfo {author} {\bibfnamefont {V.}~\bibnamefont {Gonta}}, \bibinfo
  {author} {\bibfnamefont {K.}~\bibnamefont {Krupa}}, \bibinfo {author}
  {\bibfnamefont {O.}~\bibnamefont {Sidelnikov}}, \bibinfo {author}
  {\bibfnamefont {S.}~\bibnamefont {Turitsyn}}, \bibinfo {author}
  {\bibfnamefont {M.}~\bibnamefont {Fedoruk}}, \bibinfo {author} {\bibfnamefont
  {S.}~\bibnamefont {Babin}}, \ and\ \bibinfo {author} {\bibfnamefont
  {S.}~\bibnamefont {Wabnitz}},\ }\href@noop {} {\bibfield  {journal} {\bibinfo
   {journal} {Physical review letters}\ }\textbf {\bibinfo {volume} {122}},\
  \bibinfo {pages} {103902} (\bibinfo {year} {2019})}\BibitemShut {NoStop}%
\bibitem [{\citenamefont {Liu}\ \emph {et~al.}(2016)\citenamefont {Liu},
  \citenamefont {Wright}, \citenamefont {Christodoulides},\ and\ \citenamefont
  {Wise}}]{liu2016kerr}%
  \BibitemOpen
  \bibfield  {author} {\bibinfo {author} {\bibfnamefont {Z.}~\bibnamefont
  {Liu}}, \bibinfo {author} {\bibfnamefont {L.~G.}\ \bibnamefont {Wright}},
  \bibinfo {author} {\bibfnamefont {D.~N.}\ \bibnamefont {Christodoulides}}, \
  and\ \bibinfo {author} {\bibfnamefont {F.~W.}\ \bibnamefont {Wise}},\
  }\href@noop {} {\bibfield  {journal} {\bibinfo  {journal} {Optics letters}\
  }\textbf {\bibinfo {volume} {41}},\ \bibinfo {pages} {3675} (\bibinfo {year}
  {2016})}\BibitemShut {NoStop}%
\bibitem [{\citenamefont {Krupa}\ \emph {et~al.}(2019)\citenamefont {Krupa},
  \citenamefont {Casta{\~n}eda}, \citenamefont {Tonello}, \citenamefont
  {Niang}, \citenamefont {Kharenko}, \citenamefont {Fabert}, \citenamefont
  {Couderc}, \citenamefont {Millot}, \citenamefont {Minoni}, \citenamefont
  {Modotto} \emph {et~al.}}]{krupa2019nonlinear}%
  \BibitemOpen
  \bibfield  {author} {\bibinfo {author} {\bibfnamefont {K.}~\bibnamefont
  {Krupa}}, \bibinfo {author} {\bibfnamefont {G.~G.}\ \bibnamefont
  {Casta{\~n}eda}}, \bibinfo {author} {\bibfnamefont {A.}~\bibnamefont
  {Tonello}}, \bibinfo {author} {\bibfnamefont {A.}~\bibnamefont {Niang}},
  \bibinfo {author} {\bibfnamefont {D.~S.}\ \bibnamefont {Kharenko}}, \bibinfo
  {author} {\bibfnamefont {M.}~\bibnamefont {Fabert}}, \bibinfo {author}
  {\bibfnamefont {V.}~\bibnamefont {Couderc}}, \bibinfo {author} {\bibfnamefont
  {G.}~\bibnamefont {Millot}}, \bibinfo {author} {\bibfnamefont
  {U.}~\bibnamefont {Minoni}}, \bibinfo {author} {\bibfnamefont
  {D.}~\bibnamefont {Modotto}},  \emph {et~al.},\ }\href@noop {} {\bibfield
  {journal} {\bibinfo  {journal} {Optics letters}\ }\textbf {\bibinfo {volume}
  {44}},\ \bibinfo {pages} {171} (\bibinfo {year} {2019})}\BibitemShut
  {NoStop}%
\bibitem [{\citenamefont {Wright}\ \emph {et~al.}(2016)\citenamefont {Wright},
  \citenamefont {Liu}, \citenamefont {Nolan}, \citenamefont {Li}, \citenamefont
  {Christodoulides},\ and\ \citenamefont {Wise}}]{wright2016self}%
  \BibitemOpen
  \bibfield  {author} {\bibinfo {author} {\bibfnamefont {L.~G.}\ \bibnamefont
  {Wright}}, \bibinfo {author} {\bibfnamefont {Z.}~\bibnamefont {Liu}},
  \bibinfo {author} {\bibfnamefont {D.~A.}\ \bibnamefont {Nolan}}, \bibinfo
  {author} {\bibfnamefont {M.-J.}\ \bibnamefont {Li}}, \bibinfo {author}
  {\bibfnamefont {D.~N.}\ \bibnamefont {Christodoulides}}, \ and\ \bibinfo
  {author} {\bibfnamefont {F.~W.}\ \bibnamefont {Wise}},\ }\href@noop {}
  {\bibfield  {journal} {\bibinfo  {journal} {Nature Photonics}\ }\textbf
  {\bibinfo {volume} {10}},\ \bibinfo {pages} {771} (\bibinfo {year}
  {2016})}\BibitemShut {NoStop}%
\bibitem [{\citenamefont {Guenard}\ \emph {et~al.}(2017)\citenamefont
  {Guenard}, \citenamefont {Krupa}, \citenamefont {Dupiol}, \citenamefont
  {Fabert}, \citenamefont {Bendahmane}, \citenamefont {Kerm{\`e}ne},
  \citenamefont {Desfarges-Berthelemot}, \citenamefont {Auguste}, \citenamefont
  {Tonello}, \citenamefont {Barth{\'e}l{\'e}my} \emph
  {et~al.}}]{guenard2017kerr}%
  \BibitemOpen
  \bibfield  {author} {\bibinfo {author} {\bibfnamefont {R.}~\bibnamefont
  {Guenard}}, \bibinfo {author} {\bibfnamefont {K.}~\bibnamefont {Krupa}},
  \bibinfo {author} {\bibfnamefont {R.}~\bibnamefont {Dupiol}}, \bibinfo
  {author} {\bibfnamefont {M.}~\bibnamefont {Fabert}}, \bibinfo {author}
  {\bibfnamefont {A.}~\bibnamefont {Bendahmane}}, \bibinfo {author}
  {\bibfnamefont {V.}~\bibnamefont {Kerm{\`e}ne}}, \bibinfo {author}
  {\bibfnamefont {A.}~\bibnamefont {Desfarges-Berthelemot}}, \bibinfo {author}
  {\bibfnamefont {J.-L.}\ \bibnamefont {Auguste}}, \bibinfo {author}
  {\bibfnamefont {A.}~\bibnamefont {Tonello}}, \bibinfo {author} {\bibfnamefont
  {A.}~\bibnamefont {Barth{\'e}l{\'e}my}},  \emph {et~al.},\ }\href@noop {}
  {\bibfield  {journal} {\bibinfo  {journal} {Optics Express}\ }\textbf
  {\bibinfo {volume} {25}},\ \bibinfo {pages} {4783} (\bibinfo {year}
  {2017})}\BibitemShut {NoStop}%
\bibitem [{\citenamefont {Dupiol}\ \emph {et~al.}(2018)\citenamefont {Dupiol},
  \citenamefont {Krupa}, \citenamefont {Tonello}, \citenamefont {Fabert},
  \citenamefont {Modotto}, \citenamefont {Wabnitz}, \citenamefont {Millot},\
  and\ \citenamefont {Couderc}}]{dupiol2018interplay}%
  \BibitemOpen
  \bibfield  {author} {\bibinfo {author} {\bibfnamefont {R.}~\bibnamefont
  {Dupiol}}, \bibinfo {author} {\bibfnamefont {K.}~\bibnamefont {Krupa}},
  \bibinfo {author} {\bibfnamefont {A.}~\bibnamefont {Tonello}}, \bibinfo
  {author} {\bibfnamefont {M.}~\bibnamefont {Fabert}}, \bibinfo {author}
  {\bibfnamefont {D.}~\bibnamefont {Modotto}}, \bibinfo {author} {\bibfnamefont
  {S.}~\bibnamefont {Wabnitz}}, \bibinfo {author} {\bibfnamefont
  {G.}~\bibnamefont {Millot}}, \ and\ \bibinfo {author} {\bibfnamefont
  {V.}~\bibnamefont {Couderc}},\ }\href@noop {} {\bibfield  {journal} {\bibinfo
   {journal} {Optics letters}\ }\textbf {\bibinfo {volume} {43}},\ \bibinfo
  {pages} {587} (\bibinfo {year} {2018})}\BibitemShut {NoStop}%
\bibitem [{\citenamefont {Niang}\ \emph {et~al.}(2019)\citenamefont {Niang},
  \citenamefont {Mansuryan}, \citenamefont {Krupa}, \citenamefont {Tonello},
  \citenamefont {Fabert}, \citenamefont {Leproux}, \citenamefont {Modotto},
  \citenamefont {Egorova}, \citenamefont {Levchenko}, \citenamefont {Lipatov}
  \emph {et~al.}}]{niang2019spatial}%
  \BibitemOpen
  \bibfield  {author} {\bibinfo {author} {\bibfnamefont {A.}~\bibnamefont
  {Niang}}, \bibinfo {author} {\bibfnamefont {T.}~\bibnamefont {Mansuryan}},
  \bibinfo {author} {\bibfnamefont {K.}~\bibnamefont {Krupa}}, \bibinfo
  {author} {\bibfnamefont {A.}~\bibnamefont {Tonello}}, \bibinfo {author}
  {\bibfnamefont {M.}~\bibnamefont {Fabert}}, \bibinfo {author} {\bibfnamefont
  {P.}~\bibnamefont {Leproux}}, \bibinfo {author} {\bibfnamefont
  {D.}~\bibnamefont {Modotto}}, \bibinfo {author} {\bibfnamefont
  {O.}~\bibnamefont {Egorova}}, \bibinfo {author} {\bibfnamefont
  {A.}~\bibnamefont {Levchenko}}, \bibinfo {author} {\bibfnamefont
  {D.}~\bibnamefont {Lipatov}},  \emph {et~al.},\ }\href@noop {} {\bibfield
  {journal} {\bibinfo  {journal} {Optics express}\ }\textbf {\bibinfo {volume}
  {27}},\ \bibinfo {pages} {24018} (\bibinfo {year} {2019})}\BibitemShut
  {NoStop}%
\bibitem [{\citenamefont {Leventoux}\ \emph {et~al.}(2019)\citenamefont
  {Leventoux}, \citenamefont {Parriaux}, \citenamefont {Granger}, \citenamefont
  {Jossent}, \citenamefont {Lavoute}, \citenamefont {Gaponov}, \citenamefont
  {Fabert}, \citenamefont {Tonello}, \citenamefont {Krupa}, \citenamefont
  {Desfarges-Berthelemot} \emph {et~al.}}]{leventoux2019multimode}%
  \BibitemOpen
  \bibfield  {author} {\bibinfo {author} {\bibfnamefont {Y.}~\bibnamefont
  {Leventoux}}, \bibinfo {author} {\bibfnamefont {A.}~\bibnamefont {Parriaux}},
  \bibinfo {author} {\bibfnamefont {G.}~\bibnamefont {Granger}}, \bibinfo
  {author} {\bibfnamefont {M.}~\bibnamefont {Jossent}}, \bibinfo {author}
  {\bibfnamefont {L.}~\bibnamefont {Lavoute}}, \bibinfo {author} {\bibfnamefont
  {D.}~\bibnamefont {Gaponov}}, \bibinfo {author} {\bibfnamefont
  {M.}~\bibnamefont {Fabert}}, \bibinfo {author} {\bibfnamefont
  {A.}~\bibnamefont {Tonello}}, \bibinfo {author} {\bibfnamefont
  {K.}~\bibnamefont {Krupa}}, \bibinfo {author} {\bibfnamefont
  {A.}~\bibnamefont {Desfarges-Berthelemot}},  \emph {et~al.},\ }in\ \href@noop
  {} {\emph {\bibinfo {booktitle} {2019 Conference on Lasers and Electro-Optics
  (CLEO)}}}\ (\bibinfo {organization} {IEEE},\ \bibinfo {year} {2019})\ pp.\
  \bibinfo {pages} {1--2}\BibitemShut {NoStop}%
\bibitem [{\citenamefont {Gu{\'e}nard}\ \emph {et~al.}(2019)\citenamefont
  {Gu{\'e}nard}, \citenamefont {Krupa}, \citenamefont {Tonello}, \citenamefont
  {Fabert}, \citenamefont {Auguste}, \citenamefont {Humbert}, \citenamefont
  {Leparmentier}, \citenamefont {Ducl{\`e}re}, \citenamefont {Chenu},
  \citenamefont {Delaizir} \emph {et~al.}}]{guenard2019spatial}%
  \BibitemOpen
  \bibfield  {author} {\bibinfo {author} {\bibfnamefont {R.}~\bibnamefont
  {Gu{\'e}nard}}, \bibinfo {author} {\bibfnamefont {K.}~\bibnamefont {Krupa}},
  \bibinfo {author} {\bibfnamefont {A.}~\bibnamefont {Tonello}}, \bibinfo
  {author} {\bibfnamefont {M.}~\bibnamefont {Fabert}}, \bibinfo {author}
  {\bibfnamefont {J.-L.}\ \bibnamefont {Auguste}}, \bibinfo {author}
  {\bibfnamefont {G.}~\bibnamefont {Humbert}}, \bibinfo {author} {\bibfnamefont
  {S.}~\bibnamefont {Leparmentier}}, \bibinfo {author} {\bibfnamefont {J.-R.}\
  \bibnamefont {Ducl{\`e}re}}, \bibinfo {author} {\bibfnamefont
  {S.}~\bibnamefont {Chenu}}, \bibinfo {author} {\bibfnamefont
  {G.}~\bibnamefont {Delaizir}},  \emph {et~al.},\ }\href@noop {} {\bibfield
  {journal} {\bibinfo  {journal} {Optical Fiber Technology}\ }\textbf {\bibinfo
  {volume} {53}},\ \bibinfo {pages} {102014} (\bibinfo {year}
  {2019})}\BibitemShut {NoStop}%
\bibitem [{\citenamefont {Arab{\'\i}}\ \emph {et~al.}(2018)\citenamefont
  {Arab{\'\i}}, \citenamefont {Kudlinski}, \citenamefont {Mussot},\ and\
  \citenamefont {Conforti}}]{arabi2018geometric}%
  \BibitemOpen
  \bibfield  {author} {\bibinfo {author} {\bibfnamefont {C.~M.}\ \bibnamefont
  {Arab{\'\i}}}, \bibinfo {author} {\bibfnamefont {A.}~\bibnamefont
  {Kudlinski}}, \bibinfo {author} {\bibfnamefont {A.}~\bibnamefont {Mussot}}, \
  and\ \bibinfo {author} {\bibfnamefont {M.}~\bibnamefont {Conforti}},\
  }\href@noop {} {\bibfield  {journal} {\bibinfo  {journal} {Physical Review
  A}\ }\textbf {\bibinfo {volume} {97}},\ \bibinfo {pages} {023803} (\bibinfo
  {year} {2018})}\BibitemShut {NoStop}%
\bibitem [{\citenamefont {Guasoni}(2015)}]{guasoni2015generalized}%
  \BibitemOpen
  \bibfield  {author} {\bibinfo {author} {\bibfnamefont {M.}~\bibnamefont
  {Guasoni}},\ }\href@noop {} {\bibfield  {journal} {\bibinfo  {journal}
  {Physical Review A}\ }\textbf {\bibinfo {volume} {92}},\ \bibinfo {pages}
  {033849} (\bibinfo {year} {2015})}\BibitemShut {NoStop}%
\bibitem [{\citenamefont {Mondal}\ and\ \citenamefont
  {Varshney}(2019)}]{mondal2019modal}%
  \BibitemOpen
  \bibfield  {author} {\bibinfo {author} {\bibfnamefont {P.}~\bibnamefont
  {Mondal}}\ and\ \bibinfo {author} {\bibfnamefont {S.~K.}\ \bibnamefont
  {Varshney}},\ }\href@noop {} {\bibfield  {journal} {\bibinfo  {journal}
  {Journal of Optics}\ }\textbf {\bibinfo {volume} {22}},\ \bibinfo {pages}
  {015501} (\bibinfo {year} {2019})}\BibitemShut {NoStop}%
\end{thebibliography}%

\end{document}